# Numerical Investigations of Jet A–Hexane Binary Fuel Droplet Impact on a Heated Solid Surface

Arghya Paul[1], Kanak Raj[2], Pratim Kumar*[2]

[1]*Department of Mechanical Engineering, IISc Bengaluru*
[2]*Department of Aerospace Engineering & Applied Mechanics, IIEST Shibpur*

**Abstract**—In the present work, Jet A-Hexane binary fuel droplet impact dynamics on heated solid surfaces were studied numerically. This study is crucial for practical applications such as fuel injection in combustors and thermal management of engine components. Volume of fluid (VOF) method was used to analyse the impact dynamics, spreading behaviour, vaporisation, and heat transfer of n-hexane and Jet-A blended fuel droplets on heated stainless-steel surfaces. Droplet impact dynamics were investigated for two Weber numbers, i.e., 25 and 50, and surface temperatures ranging from 50 °C to 227 °C to capture transitions from gentle spreading to nucleate boiling and rebound phenomena. This work examines how fuel blending influences inertia, lamella formation, vapour recoil, and film boiling regimes. The results show that higher inertia in blended fuels enhances spreading but also triggers stronger vapour recoil at elevated temperatures, leading to droplet rebound. In contrast, pure hexane transitions to a stable film boiling regime at high surface temperatures, resulting in a decline in smoother heat flux. New correlations were developed linking Weber number, spreading ratio, and wall heat flux, offering predictive insights for real-world combustion scenarios. These findings advance the understanding of bi-component fuel droplet impacts on heated surfaces and provide a framework for designing efficient spray systems in combustors and thermal management in propulsion and power generation applications.

*corresponding author e-mail: pratim.kumar.86@gmail.com*

## 1. INTRODUCTION

The study of droplets impacting surfaces has a long history, dating back to Worthington's early works [1]. His observations of splashing and jetting phenomena opened diverse areas of research in dynamics of droplet impact. The application of this study is very vast, such as fire suppression with sprinkler systems, evaporative cooling of electronic systems by droplet streams, spray injection in confined combustion chambers, and rain droplet impingement onto the gas turbine blades during in-flight rain ingestion. The size of droplets can vary from millimeters to micrometers scale, and the impinging surfaces can also of different properties, such as heated or normal temperature, different materials, different roughness's, and flat or curved surface. Out of all these applications, our study is relevant to the combustion of spray in a combustion chamber. When a stream of spray is injected, it interacts with the wall and undergoes numerous processes. So, to understand those processes, we can rely on the results obtained from a single droplet impacting a heated surface [2–5].

The interaction between fuel spray droplets and heated solid surfaces is a critical phenomenon in engine combustors, where droplets impinge upon high-temperature components such as combustion chamber walls, piston crowns, or turbine blades. This interaction governs key processes such as heat transfer, droplet evaporation, wall wetting, and secondary atomization, all of which strongly influence combustion efficiency, pollutant formation, and the distribution of thermal loads within the engine system [6]. The outcome of fuel spray impingement is governed by a multitude of factors, including spray characteristics (e.g., droplet size, velocity, number flux, and spatial distribution), liquid physical properties (density ($\rho$), surface tension ($\sigma$), and dynamic viscosity($\mu$)), surface attributes (such as roughness and chemical composition), and thermal conditions (temperatures of the liquid, surface, and ambient environment), as well as ambient pressure [4]. The inherent complexity of spray impingement phenomena is further amplified by various interrelated sub-processes, including droplet-droplet interactions in the free spray and on the surface, the dynamic evolution of thin liquid films, interactions between incoming droplets and pre-existing films, secondary droplet formation due to film disturbances, and cascading interactions among both primary and secondary droplets [7–10].

When a liquid droplet impinges on an isothermal flat solid surface, its motion driven by inertia and gravitational forces initiates the spreading stage, during which the droplet flattens and extends radially across the surface. As spreading progresses, capillary (surface tension) and viscous forces resist further motion, gradually slowing the advancing contact line. Eventually, these forces balance the initial momentum, causing the liquid–gas interface to cease outward motion and reach a maximum spread diameter. Following this, surface tension acts to retract the interface inward, initiating the recoiling phase, wherein the droplet may partially retract or even rebound, depending on the surface and fluid properties. At elevated surface temperatures, evaporation becomes significant, further complicating the dynamics of spreading and recoiling by altering interfacial tension, internal flow fields, and mass loss. The combined phenomena of droplet impact, spreading, recoiling, and evaporation play a crucial role in various applications, including spray cooling, fuel injection, and surface coating. Notably, the temperature of the heated solid surface ($T_s$) critically influences these processes, as it governs the dominant heat transfer mechanisms, conduction, convection, or phase change, between the droplet and the substrate. Several studies [1,2,11–14] have examined the morphological dynamics of liquid drops at various heat transfer regimes, including film boiling ($T_s > T_L$), nucleate boiling ($T_b < T_s < T_L$), and film evaporation ($T_\infty \sim T_s < T_b$). $T_L$ is the Leidenfrost point, $T_b$ is the boiling point of the drop liquid at ambient pressure, $T_\infty$ is the ambient temperature, and $T_s$ is the surface temperature. Chandra and Avedisian [1] investigated how Ts affects the spreading and retreating of a liquid drop upon impact.

To deepen the understanding of solid–liquid–gas interfacial phenomena, several researchers have conducted experimental investigations focused on droplet dynamics and heat transfer. Crafton and Black [15] studied water and heptane droplet evaporation on heated horizontal surfaces, demonstrating significant differences in wetted diameter, height, and contact angle based on liquid properties. Bhardwaj et al. [16] measured interfacial temperatures using laser thermo-reflectance, revealing an initial exponential rise followed by linear cooling, highlighting coupled heat and mass transfer during droplet evaporation. Chatzikyriakou et al. [17] developed a method to quantify heat transfer from rebounding non-wetting droplets, enabling direct measurement of energy extracted during impact. Bhat et al. [18] investigated film evaporation for various fuels (heptane, decane, Jet A-1, diesel) on heated stainless steel, showing that the maximum spread factor decreases with normalised surface temperature at low Weber numbers, but increases at higher values due to viscosity effects.

Sen et al. [19] studied camelina-based biofuel droplets across a wide range of Weber numbers and temperatures, finding the dynamic Leidenfrost point to be largely Weber-independent and closely aligned with thermodynamic predictions. Zhang et al. [20] and Boettcher et al. [21] examined ignition characteristics of n-hexane, reporting strong sensitivity to heating rate, equivalence ratio, and composition, with scenarios permitting complete vaporisation without ignition. Kompinsky et al. [22] explored binary fuel mixtures (n-hexane/n-decane) and identified six impact regimes, noting that higher impact velocity reduced the surface temperature threshold for regime transitions. Kitano et al. [23] observed complex, composition-dependent evaporation patterns in surrogate Jet-A fuels. Chausalkar et al. [24] systematically analysed bi- and multi-component fuels, including commercial blends, revealing that increasing the volatile fraction enhances film breakup, decreases mean droplet size, and shifts regime transitions to lower temperatures.

Numerical simulations have become an effective and cost-efficient tool for exploring droplet impact and evaporation phenomena, offering detailed insights into interfacial dynamics that are often difficult to capture experimentally. For instance, Nikolopoulos et al. [25] modelled n-heptane and water droplets impacting heated surfaces under sub and super-Leidenfrost regimes using a finite volume method with the VOF approach, successfully capturing droplet morphology, evaporation rates, and temperature and vapour fields. Building on this, Strotos et al. [26] applied adaptive meshing to simulate mono and bi-component droplets with varying compositions, resolving coupled flow, thermal, and mass transfer interactions. Banerjee [27] investigated ethanol-iso-octane droplet evaporation numerically, showing how ambient and fluid conditions influence evaporation characteristics. Other studies using dynamic contact angle and VOF models have demonstrated how Weber number and groove geometry control anisotropic spreading and Cassie–Wenzel transitions on grooved superhydrophobic surfaces, providing guidance for anti-icing and microfluidic designs [28].

Research on droplet impact on PMMA and PTFE surfaces with varying roughness and surfactant use (Weber numbers 20–300) has shown how these factors affect spreading, oscillations, and rebound behaviour, leading to predictive models for maximum spreading with good accuracy (<5% deviation) [29]. In molten tin droplet impacts, increasing impact velocity and substrate temperature intensified splashing, but did not always raise the spreading factor instead, solidification and surface energy changes dominated fragmentation as temperatures rose from 50 °C to 290 °C [30]. Boiling simulations of $Al_2O_3/H_2O$ and $CuO/H_2O$ nanofluids in copper metal foam pipes revealed that lower porosity ($\varepsilon = 0.80$) and lower pore density (5 PPI) improved heat transfer performance with manageable pressure drop, and the updated model predicted HTC up to 7% more accurately than prior methods [31]. Studies using the CLSVOF method found that droplet impact angle, spacing, and liquid film thickness strongly influence interface shape, splash temperature, and wall heat transfer, with perpendicular impacts achieving heat transfer coefficients up to 9903 $W/m^2 \cdot K$ and thin films enhancing local peaks through stronger convection [32]. Additionally, protrusion size on textured surfaces affects droplet-induced heat transfer by promoting local mixing but may reduce overall plane area cooling [33]. Finally, customised ANSYS FLUENT simulations with a sharp interface VOF and adaptive mesh refinement captured 3D nucleate boiling behaviour efficiently, matching experimental results while significantly reducing computational cost, and showing promise for complex boiling systems with multiple nucleation sites [34].

Although numerous studies have examined the dynamics of fuel droplet impingement on heated surfaces, including spreading, levitation, breakup, and evaporation, most have focused on single-component fuels or binary blends with closely matched boiling characteristics. There remains limited understanding of how small additions of volatile components, such as 10% n-hexane in Jet fuel, influence the onset and evolution of boiling regimes, spreading behaviour, and droplet breakup across a range of surface temperatures. Specifically, the shift in nucleate boiling onset from 120 °C and 180 °C for pure hexane to 227 °C for the blend indicates a suppression of phase change dynamics caused by the dominant low-volatility base fuel, an effect not well captured in existing numerical models. To address this gap, this paper develops a validated numerical model using the Volume of Fluid (VOF) method to simulate droplet impingement and evaporation. The model incorporates Blake's molecular-kinetics-based dynamic contact angle formulation for contact line motion and Schrage's evaporation model to capture interfacial mass transfer, while a species transport equation is solved to describe vapour diffusion and convection. The model is used to compare the impact dynamics of pure n-hexane and 10% hexane–90% Jet fuel blend droplets at Weber numbers 25 and 50 across various surface temperatures, with the goal of identifying regime boundaries and quantifying how minor volatile additions affect spreading, breakup, and evaporation behaviour in evaporation-relevant combustion conditions.

## 2. COMPUTATIONAL MODEL

The current investigation employs a two-dimensional rectangular computational domain measuring 60 mm in width and 30 mm in height, as shown in Figure 1. The domain includes a single droplet with an initial diameter positioned near the heated wall surface. The bottom boundary was maintained at a constant wall temperature, while the side and top boundaries were set as pressure outlets. The specific boundary conditions like droplet diameter, Weber number, surface temperature, initial droplet temperature, operating pressure, and surface roughness are summarised in Table 1, while the thermophysical properties of Hexane, Jet A, and their blend are provided in Table 2. For fluid properties, thermodynamic and transport parameters were treated as temperature and species-dependent. Specifically, physical property data for hexane were obtained from the NIST database [35] and fitted using piecewise polynomial equations, while material properties for Jet A were taken directly from the ANSYS Fluent material database. Mixture density, specific heat, viscosity, and thermal conductivity for the blend were calculated using volume-weighted and mass-weighted mixing laws. The Volume of Fluid (VOF) approach was employed to capture the liquid–vapour interface, and the Courant number was kept below 1 with a time step size of $1 \times 10^{-6}$ s to ensure numerical stability during strong evaporation-induced pressure and velocity changes.

### 2.1 Governing Equation

The governing VOF equation, shown in Equation (1), defines the volume fraction $\alpha_q$ for the $q^{th}$ phase as the ratio of its volume $V_q$ to the total cell volume, ensuring that the sum of volume fractions equals one within each control volume.

$$\alpha_q = \frac{V_q}{\sum V_q} \tag{1}$$

Several conservative equations govern the fluid flow, heat transfer, and vapour distribution in the system, requiring a complete and accurate discretised solution to resolve the underlying physics. In this context, the VOF equation is employed to track the evolution of the phase interfaces within the computational domain, ensuring precise capture of the droplet–vapour interactions and is presented as Equation 2.

$$\frac{1}{\rho_q}[\frac{\partial(\alpha_q \rho_q)}{\partial t} + \nabla \cdot (\alpha_q \rho_q \vec{v_q})] = S_{\alpha_q} \tag{2}$$

There are two different phases in this analysis: liquid and gas. The VOF approach determines the volume fraction ($\alpha_l$) of the liquid phase in each computing cell and subtracts this number from one to determine the gas phase fraction. In order to guarantee that the transport parameters, including density, are assessed as volume-averaged values based on the contributions of each phase, evaporation at the liquid–gas interface generates a source term ($S_{\alpha_q}$) that alters the local volume fraction during phase change and is given as Equation 3.

$$\rho = \sum_{q=1}^{2} \alpha_q \rho_q = \alpha_l \rho_l + (1-\alpha_l)\rho_g = 1 \tag{3}$$

The two-dimensional unsteady conservative continuity equation is expressed as follows:

$$\frac{\partial \rho}{\partial t} + \nabla \cdot (\rho \vec{v}) = 0 \tag{4}$$

where $\vec{v}$ is the velocity vector of fluid. The mass conservation equation for fluid flow is also represented by this equation.

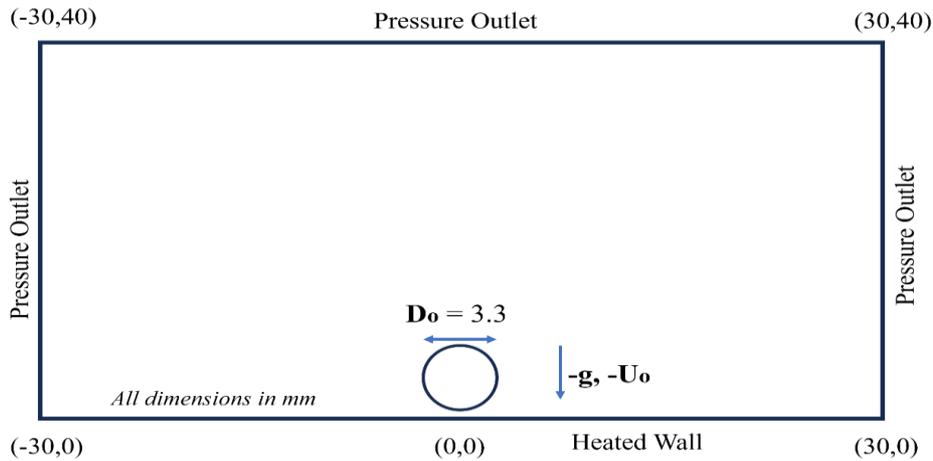

**Figure 1:** Schematic diagram of a single droplet impacting on the heated surface along with boundary conditions.

**Table 1:** Boundary conditions used in the current investigation.

| | |
|---|---|
| Droplet Diameter (mm) | 3.3 |
| Weber Number | 25, 50 |
| Surface Temperature (K) | 323, 393, 453, 500 |
| Initial Droplet Temperature (K) | 300 |
| Operating Pressure (bar) | 1 |
| Surface Roughness (μm) | 3.75 |

**Table 2:** Thermophysical properties used in the current investigation.

| Properties | Hexane | Jet A | Blend |
|---|---|---|---|
| Density (ρ) (kg/m3) | $\rho = 851.5028 + (-0.052)T + (-0.004)T^2 + (1.11863 \times 10^{-5})T^3 + (-1.19998 \times 10^{-8})T^4$ | 803 [36] | - |
| Specific Heat Capacity ($C_p$) (J/Kg.K) | $C_p = 4042.3429 + (-28.0657)T + 0.1308T^2 + (-0.0002)T^3 + (1.8243 \times 10^{-7})T^4$ | 2066.75 | - |
| Thermal Conductivity (K) (W/m.K) | $K = 0.338 + (-0.0016)T + (5.012 \times 10^{-6})T^2 + (-9.18 \times 10^{-9})T^3 + (7.0886 \times 10^{-12})T^4$ | 0.1187 | - |
| Viscosity (μ) (Pa.s) | $\mu = 0.0132 + (-0.0001)T + (5.7464 \times 10^{-7})T^2 + (-1.0901 \times 10^{-9})T^3 + (7.8984 \times 10^{-13})T^4$ | 0.0012676 | 0.001129 (Mixing Law) [37] |
| Surface Tension (σ) (N/m) | 0.017897 | - | 0.02686 [38] |
| Boiling Point (K) | 342 | 483 [36] | - |

The momentum equation used in this analysis is expressed below as Equation 5:

$$\frac{\partial(\rho \vec{v})}{\partial t} + \nabla \cdot (\rho \vec{v}\vec{v}) = -\nabla P + \nabla \cdot [\mu(\nabla \vec{v} + \nabla \vec{v}^T)] + \rho \vec{g} + \vec{F_\sigma} \quad (5)$$

Here, P represents the pressure and $\vec{v}$ denotes the velocity vector. The term $\vec{F_\sigma}$ is responsible for the surface tension force vector, which is non-zero only at the liquid–gas interface. This force vector must be defined carefully, as it is essential for capturing the impingement dynamics accurately. Droplet spreading, deformation, and internal circulation during wall impact depend on this force. The force vector $\vec{F_\sigma}$ is calculated using the capillary pressure ($P_n$) in the normal direction and the tangential pressure ($P_t$) along the droplet surface, as shown in Figure 2, and the expressions are presented as Equation 6 and Equation 7:

$$P_n = \sigma \cdot \kappa \quad (6)$$

where $\sigma, \kappa$ are surface tension coefficient and interfacial curvature respectively.

$$P_t = \nabla_t \sigma \quad (7)$$

$\vec{F_\sigma}$ can be calculated with the help of Continuous Surface Force (CSF) model given by Brackbill et al. [39] Hence, $\vec{F_\sigma}$ is given by the expression presented as Equation 8:

$$\vec{F_\sigma} = \frac{|\nabla \alpha_l| \rho}{0.5(\rho_l + \rho_g)}(\sigma \kappa n + \nabla_t \sigma) \qquad (8)$$

Where $n$ is the unit surface normal. It is defined as the ratio of surface normal vector ($\vec{n}$) and magnitude of surface normal vector. Surface normal vector ($\vec{n}$) can be calculated as gradient of volume fraction, as shown in Equations 9, 10, and 11:

$$n = \frac{\vec{n}}{|\vec{n}|} \qquad (9)$$

$$\vec{n} = \nabla \alpha_l \qquad (10)$$

$$\kappa = \nabla \cdot n \qquad (11)$$

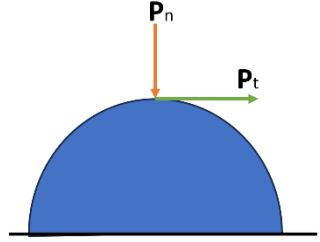

**Figure 2:** Schematic representation of body forces at the liquid–gas interface, illustrating the normal component of capillary pressure (*Pn*) and the tangential component (*Pt*) acting along the droplet surface.

Therefore, the effect of heat transfer is considered using the energy equation, which is presented as Equation 12:

$$\frac{\partial(\rho E)}{\partial t} + \nabla \cdot (\vec{v}(\rho E + p)) = \nabla \cdot (\lambda \nabla T) + S_E \qquad (12)$$

Where $S_E$ is the energy source term. It is activated because of heat production or heat loss at the interface due to the phase change process. $E$ is the energy and is represented by the expression given in Equation 13:

$$E = \frac{\sum_{q=1}^{n} \alpha_q \rho_q E_q}{\sum_{q=1}^{n} \alpha_q \rho_q} \qquad (13)$$

Turbulence plays a significant role in accurately predicting flow behaviour in droplet impact and heat transfer problems. It influences mixing, momentum exchange, and local flow instabilities near the liquid–gas interface. The standard $k$–$\varepsilon$ turbulence model was used for the numerical analysis. This model, originally proposed by Launder and Spalding [40,41], is widely accepted due to its reliability, cost-effectiveness, and accuracy [42]. All constant values were retained at their default settings. The turbulent viscosity in this model is calculated using the following expression, presented as Equation 14:

$$\mu_{T,l} = \rho_l C_\mu \frac{k^2}{\varepsilon} \tag{14}$$

The conservation expressions for the turbulent kinetic energy ($k$) and dissipative energy ($\varepsilon$) are given by the expressions presented as Equations 15 and 16:

$$\frac{\partial(\alpha_l \rho_l k)}{\partial t} + \nabla \cdot (\alpha_l \rho_l v_l k) = \nabla \cdot [\alpha_l (\mu_{L,l} + \frac{\mu_{T,l}}{\sigma_k}) \nabla k] + \alpha_l (G_k - \rho_l \varepsilon) \tag{15}$$

$$\frac{\partial(\alpha_l \rho_l \varepsilon)}{\partial t} + \nabla \cdot (\alpha_l \rho_l v_l \varepsilon) = \nabla \cdot [\alpha_l (\mu_{L,l} + \frac{\mu_{T,l}}{\sigma_\varepsilon}) \nabla \varepsilon] + \alpha_l (C_{1\varepsilon} \frac{\varepsilon}{k} G_k - C_{2\varepsilon} \rho_l \frac{\varepsilon^2}{k}) \tag{16}$$

In the above expressions, the $G_k$ is the production of the turbulent kinetic energy ($k$), which is defined by the following expression presented as Equation 17:

$$G_k = \mu_{T,l} [\nabla v_k + (\nabla v_k)^T] \cdot \nabla v_l \tag{17}$$

where $\sigma_k$ and $\sigma_G$ are the turbulent Prandtl numbers for $k$ and $\varepsilon$, respectively, $C_1$ and $C_2$ are constants.

Species transport was considered in this study to model the vaporisation of the liquid droplet into the surrounding ambient environment. It was assumed that the gas phase consists of both the vapour generated by evaporation and the surrounding air. The liquid phase transitions to the vapour phase at the liquid–gas interface through evaporation, and the produced vapour then disperses into the air by diffusion and convection. Since the evaporation rate is strongly linked to the vapour mass transfer, which is governed by both diffusion and convection, a species transport equation was solved for the gas phase to accurately capture this behaviour. The expression is presented as Equation 18:

$$\frac{\partial(\rho_q \alpha_q y_q^i)}{\partial t} + \nabla \cdot (\rho_q \alpha_q \vec{v} y_q^i) = \nabla \cdot (\rho_q \alpha_q D_{12} \nabla y_q^i) + S_i \tag{18}$$

$$D_{12} = \frac{10^{-3} T^{1.75} (\frac{1}{M_1} + \frac{1}{M_2})^{0.5}}{P[(\sum_i V_{i1})^{\frac{1}{3}} + (\sum_i V_{i2})^{\frac{1}{3}}]^2} \tag{19}$$

Where mass fraction is represented by $y$. $D_{12}$ is binary diffusion coefficient as given in Equation 19. $T$ is temperature, $M$ is molecular weight, $V_i$ is spatial diffusion parameter [49].

Accurately capturing vaporisation through the governing equations is a crucial aspect of this numerical model. Traditionally, the evaporation rate is estimated using Fick's law by several researchers [50,44], which assumes that the interface vapour concentration is at saturation to calculate the local vapour concentration gradient. However, this assumption does not hold true for the present case. This is because the high latent heat associated with the liquid–vapour phase change results in significant evaporative cooling, and the surface temperature is often near or even above the liquid's saturation temperature. Therefore, when a droplet impinges on a heated surface for cooling, strong evaporation occurs at the liquid–gas interface, driving the system into a non-equilibrium state. To address this, the molecular kinetic theory is applied to calculate the evaporating mass flux at the liquid–gas interface, $J_t$ as suggested by [43], and is presented as Equation 19:

$$J_t = \beta \sqrt{\frac{M}{2\pi R}} \left(\frac{P_{sat}}{\sqrt{T_{sat}}} - \frac{P_v}{\sqrt{T_g}}\right) \tag{19}$$

where $\beta$ is accommodation coefficient, $P_{sat}$ is saturation pressure at $T$, $P_v$ is vapor partial pressure at cell. VOF model assumes $T_l = T_g = T$ (cell temperature). Hence evaporation rate per unit cell ($m'$) can be calculated using the expression presented as Equation 20:

$$m' = |\nabla \alpha_l| J_t \tag{20}$$

This equation was turned on by clicking on the check box named 'Hertz Knudsen' in the commercial CFD software package.

**2.2 Contact Angle Dynamics**

Wetting occurs when a liquid droplet comes into contact with and spreads across a solid surface. Wetting is used to describe surfaces and evaluate solid–liquid interactions. Contact angle (θ), measures wettability. The contact angle between the solid surface and the liquid gas interface defines the contact line boundary condition for forecasting droplet flow dynamics, such as droplet form, in droplet impact simulations. The static contact angle is the angle formed when a droplet is at rest on a solid surface and in equilibrium. When the droplet is not at rest or the contact line moves, the contact angle changes to a dynamic angle. Measurements have revealed that the dynamic contact angle can differ greatly from the static contact angle. The experimentally observed dynamic contact angle, $θ_d$, is typically found to depend on both the speed and direction of the wetting line movement, implying that $θ_d$ is velocity dependent [44].

The dynamic contact angle is modelled as a function of contact line velocity. Several empirical equations exist to express the relationship between dynamic contact angle and contact line velocity [45–47]. The majority of them follow the Hoffman-Tanner-Voinov law or a broader form developed by Cox [48]. All models predict similar values for the dynamic contact angle. This study uses Kistler's correlation [44] to compute the dynamic contact angle at each time step, as presented in Equation 21:

$$\theta_d = f_H(Ca + f_H^{-1}(\theta_e)) \tag{21}$$

$$f_H = \arccos\left[1 - 2\tanh\left(5.16\left(\frac{x}{1+1.31x^{0.99}}\right)^{0.706}\right)\right] \tag{22}$$

$$Ca = \frac{\mu V_{cl}}{\sigma} \tag{23}$$

Where, *Ca* is capillary number as given in Equation 23, $\sigma$ is surface tension, $\mu$ is viscosity, $V_{cl}$ is contact line velocity, $f_H$ Hoffmann function $f_H^{-1}$ is inverse Hoffmann function, $\theta_e$ is equilibrium contact angle, $\theta_d$ is dynamic contact angle.

The dynamic contact angle (DCA) equation was implemented using a User Defined Function (UDF). The original DCA UDF was provided by Miller et al [49]. We conducted several tests with the original UDF; however, it did not perform satisfactorily for our case. Therefore, we developed an in-house UDF by modifying the original version to suit our requirements. Commercial CFD software ANSYS Fluent was used to carry out the numerical analysis of a droplet impinging on a heated flat plate. To solve the problem numerically, the Geo-Reconstruct scheme was employed to discretise the VOF equation. The Least Squares Cell-Based gradient evaluation method was selected for computing gradients and derivatives of the governing equations due to its accuracy and computational efficiency. The transport equations were discretised in time using the first-order upwind scheme, and the first-order implicit method was applied for time integration. For pressure discretisation, the PRESTO scheme was used, while pressure–velocity coupling was handled using the SIMPLE algorithm.

## 3. MESH INDEPENDENCE STUDY AND VALIDATION

A mesh independence study was performed to ensure that the mesh resolution does not influence the accuracy of the numerical results. Three different mesh densities were tested with approximately 90,000, 180,000, and 270,000 elements to evaluate the sensitivity of the droplet spreading behaviour to the discretisation size. The spreading ratio (D/D$_0$) was compared over dimensionless time for each mesh and validated against experimental data, as shown in Figure 3. Results indicate that the coarser mesh with 90,000 elements underpredicts the spreading ratio, especially during the maximum spreading phase, while the solutions obtained using 180,000 and 270,000 elements closely match the experimental trend of Deendarlianto et al. [50] throughout the entire spreading and receding stages. These results demonstrate that further mesh refinement beyond 180,000 elements yields negligible improvement. Therefore, a mesh with 180,000 elements was selected for all subsequent simulations to ensure both accuracy and computational efficiency.

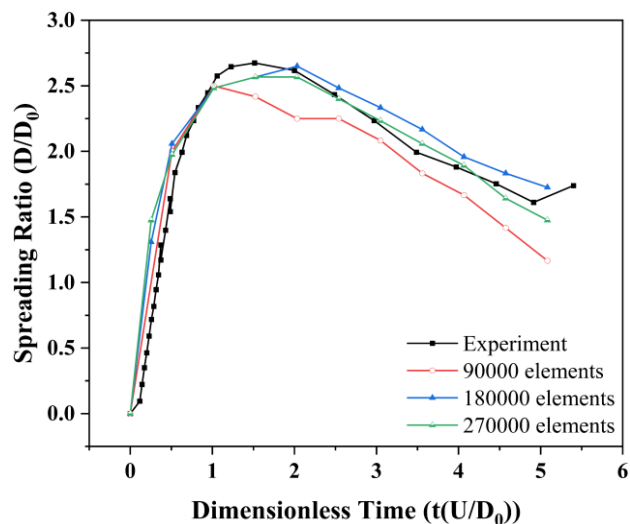

**Figure 3:** Mesh independence study result for water droplet diameter of 2.4 mm and Weber number 52.1 on heated stainless-steel surface having temperature of 433K with static contact angle of 85.7°

The developed computational model was validated against published experimental results before being applied to study droplet impact dynamics. Deendarlianto et al. [50] investigated water droplets with a diameter of 2.4 mm impinging on a practical stainless-steel flat plate (SUS 304) under varying Weber numbers and surface temperatures. For the validation, a Weber number of 52.1 and a surface temperature of 433K were selected, with a reported static contact angle of 85.7°. The model predictions for the maximum spreading ratio were compared to the experimental observations and showed good agreement, with less than 10 % deviation. To ensure the model's reliability, both quantitative and qualitative comparisons were conducted, including a frame-by-frame analysis of droplet shape evolution shown in Figure 4(a) and the temporal variation of the spreading ratio illustrated in Figure 4(b). This strong agreement confirms that the present computational model is robust and suitable for investigating droplet impact dynamics under varying conditions.

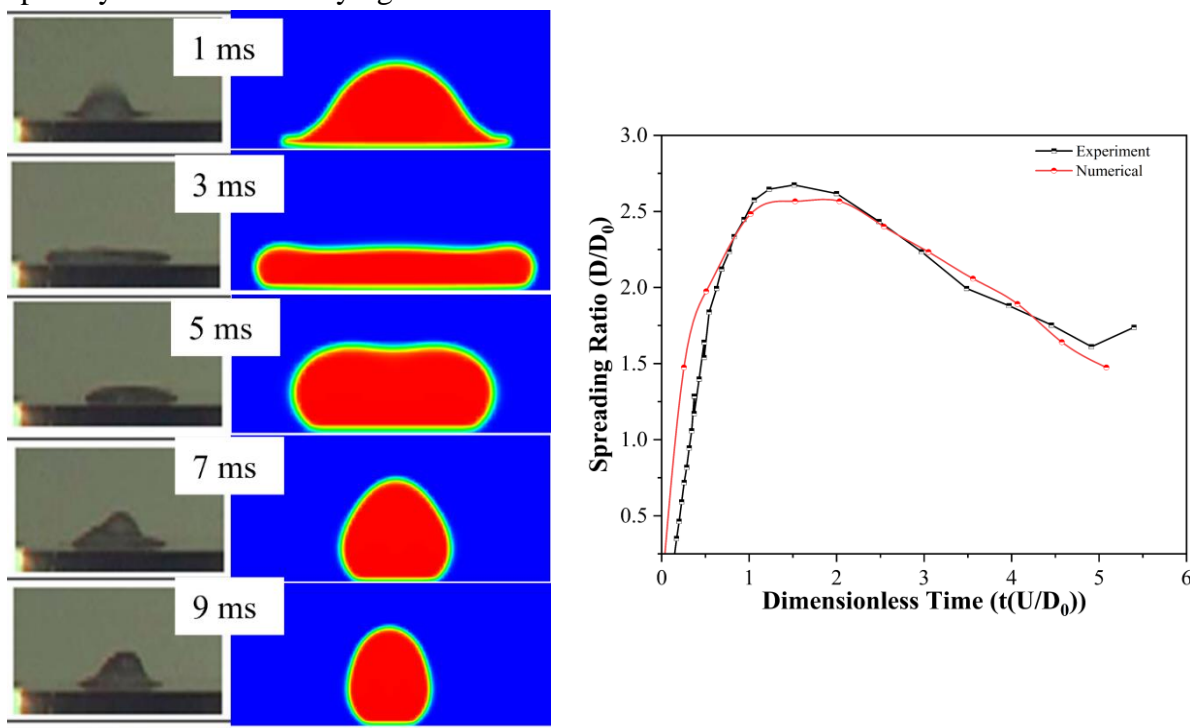

**Figure 4:** (a) Frame-by-frame comparison of experimental and numerical droplet shapes at different time instances during impact; (b) Temporal evolution of the spreading ratio ($D/D_0$) comparing experimental data with numerical predictions.

## 4. RESULTS AND DISCUSSION

This section discusses droplet morphology and impact behaviour under various thermal and dynamic conditions. Impacts are analysed for three different droplet velocities, which correspond to two Weber numbers due to differences in fuel properties between the pure and blended fuels. Blended fuel impacts are examined at Weber numbers 25 and 50, while pure hexane is evaluated at Weber number 25 for direct comparison. Surface temperature effects are studied at four levels: 323K, 393K, 453K and 500K. Comparisons illustrate how impact velocity, surface temperature, and fuel composition influence spreading, splashing, heat transfer, and overall droplet dynamics. These insights help to understand practical fuel spray behaviour on heated surfaces under realistic operating conditions.

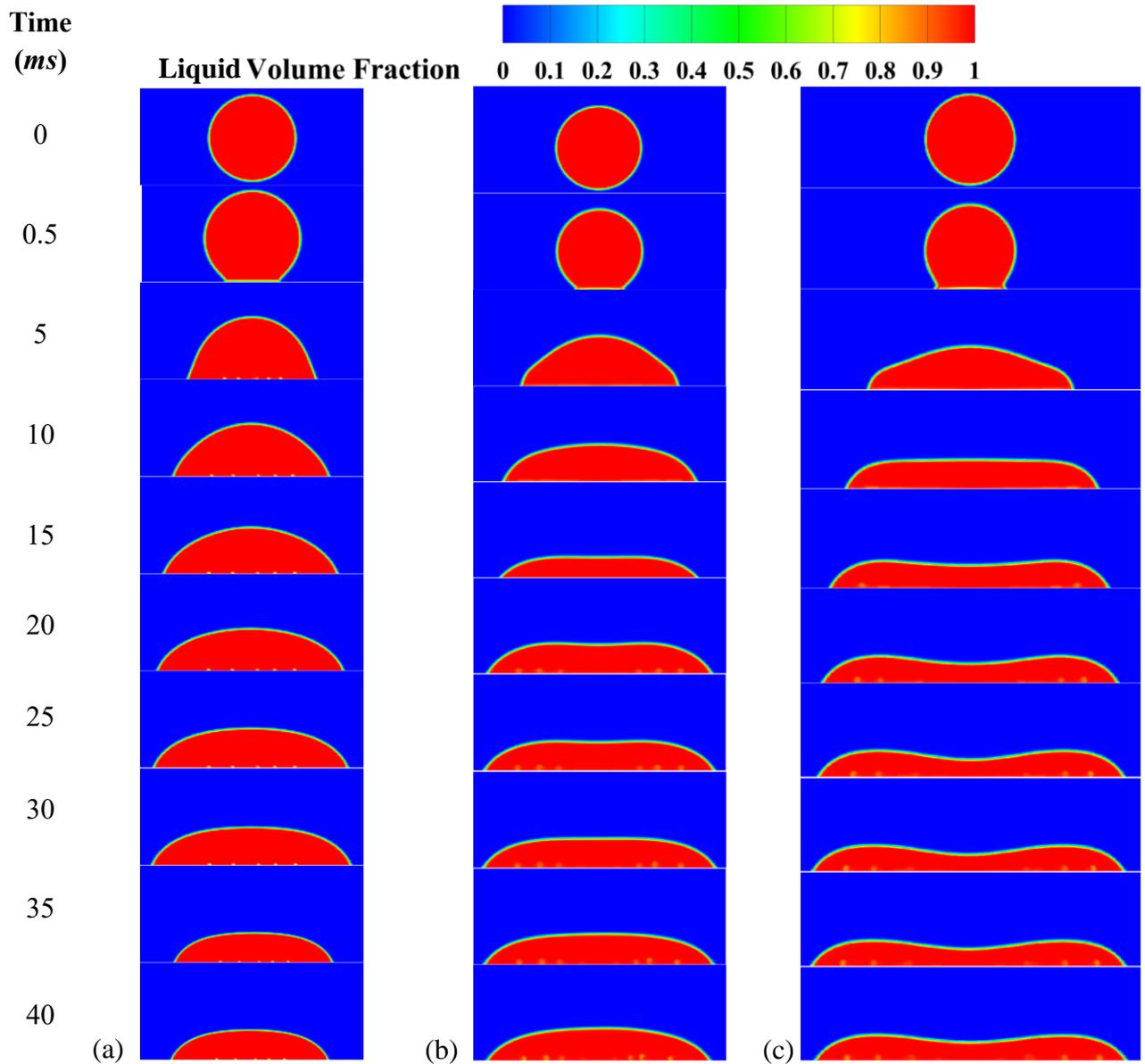

**Figure 5:** Temporal evolution of droplet morphology for (a) pure hexane at Weber number 25, (b) blended fuel (90% Jet A + 10% hexane) at Weber number 25, and (c) blended fuel at Weber number 50 during impact on a stainless-steel surface maintained at 323K.

## 4.1. Droplet Dynamics Studies

4.1.1 Droplet Impact Dynamics and Heat Transfer Behavior at 323K

At a surface temperature of 323K, the impact behaviour of a 3.3 mm droplet on a stainless-steel substrate was investigated at two Weber numbers: 25 and 50. The droplet spreading characteristics are illustrated in Figure 5. Figure 5(a) shows the spreading of pure n-hexane at Weber number 25, while Figure 5(b) and Figure 5(c) correspond to the blended fuel at Weber numbers 25 (Blend-25) and 50 (Blend-50), respectively. Although pure hexane and Blend-25 have the same Weber number, Blend-25 exhibits a larger spreading diameter. This is attributed to its higher inertia, which enables more extensive radial spreading upon impact. The increased inertia of Blend-25 arises from its higher density compared to pure n-hexane. Composed of 90% Jet A and 10% n-hexane by volume, the blend inherits the greater density of Jet A. At a constant Weber number, this implies a higher impact velocity for Blend-25, resulting in increased kinetic energy that enhances spreading. Despite its higher viscosity, Blend-25 spreads more than hexane, indicating that inertial forces dominate over viscous resistance in this regime. Similarly, the influence of surface tension is suppressed by the stronger inertial effects.

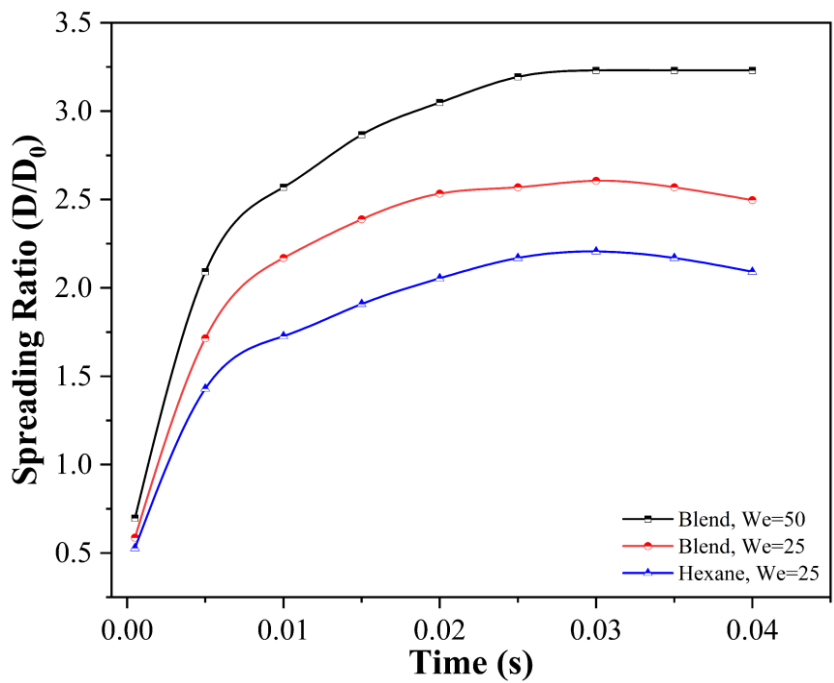

**Figure 6:** Temporal variation of the spreading ratio (D/D$_0$) for pure hexane at Weber number 25 and blended fuel (90% Jet A + 10% hexane) at Weber numbers 25 and 50 during droplet impact on a heated surface at 323K .

Lamella formation is clearly visible for Blend-25 in Figure 5(b), emerging around 20 ms after impact as the droplet spreads outward. This indicates that a portion of the kinetic energy contributes to forming a thin liquid sheet along the periphery. The lamella becomes even more prominent for Blend-50, as shown in Figure 5(c), where increased impact energy and inertia generate a stronger radial flow and a more pronounced lamella. Air entrapment near the lamella edge, as reported in previous studies [51], is also noticeable. Figure 6 provides a comparative analysis of the temporal evolution of the spreading ratio for the three cases. Pure hexane reaches its maximum spreading diameter first, followed by Blend-25 and then Blend-50. This progression highlights the role of increasing inertia and impact velocity in delaying the peak spread and shaping lamella dynamics. These results underscore how fuel composition, impact energy, and interfacial forces influence droplet spreading on heated surfaces.

The heat transfer behaviour during droplet impingement is governed by the thermal interactions at the droplet–wall interface and the fluid's internal transport properties. Figure 7 shows the average wall heat flux and wall adjacent temperature for surface temperature 323K. When a droplet at 300 K contacts a wall heated to 323 K, initial heat transfer occurs via conduction at the wall–liquid interface. This establishes a temperature gradient that triggers internal convection, especially in low-viscosity fluids [52]. The resulting convective flow distributes thermal energy throughout the droplet, enhancing its overall temperature rise. The efficiency of this combined conduction–convection mechanism is influenced by the liquid's thermal conductivity, specific heat, viscosity, and spreading behaviour. In this study, hexane exhibited the highest average heat flux, as shown in Figure 7(a). Its relatively high thermal conductivity (0.137 W/m·K) and specific heat (~2500 J/kg·K), along with low viscosity, facilitated efficient conduction from the wall and strong internal convection. Despite a comparatively smaller spreading area, the thermal transport within the droplet was highly effective.

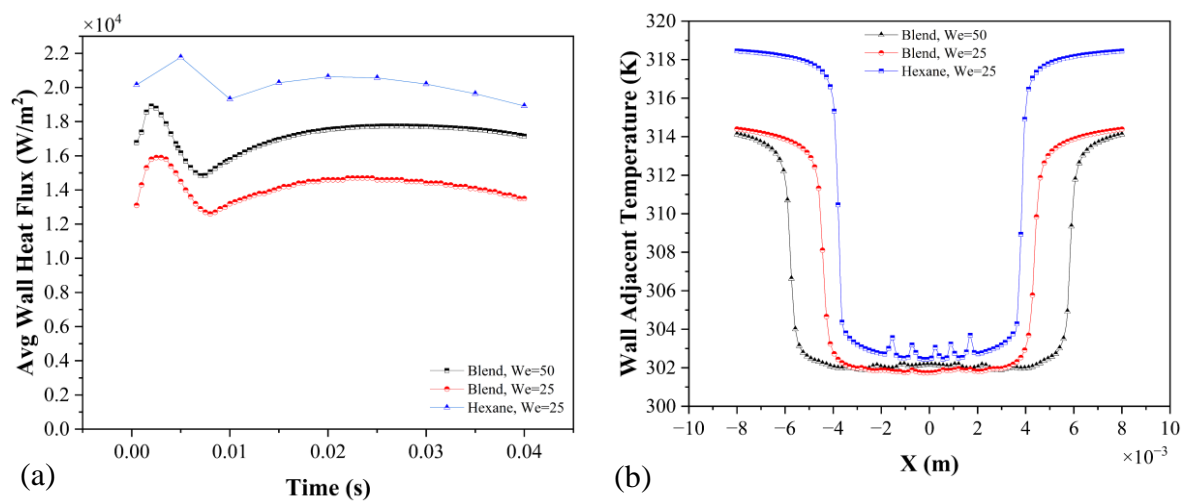

**Figure 7:** (a) Temporal variation of average wall heat flux and (b) spatial distribution of wall-adjacent temperature at 0.1 mm above the surface for pure hexane at Weber number 25 and blended fuel (90% Jet A + 10% hexane) at Weber numbers 25 and 50, at a substrate temperature of 323K .

Blend-25 showed the lowest heat flux due to its lower thermal conductivity (~0.1187 W/m·K), reduced specific heat (~2066 J/kg·K), and higher viscosity, all of which limited conduction and suppressed internal circulation. Its smaller spreading diameter and thicker film further restricted effective contact with the wall, thereby reducing heat transfer. Blend-50 yielded a moderate heat flux, higher than Blend-25 but lower than hexane. The increased Weber number provided higher impact inertia, which promoted wider spreading and thinner film formation, thereby improving wall contact. This partially compensated for the blend's weaker thermal properties. Figure 7(b) presents the temperature distribution at 0.1 mm above the wall surface, which remained nearly identical for all three cases. However, the slightly higher thermal conductivity of hexane led to an elevated heat flux. Since heat flux is directly proportional to thermal conductivity under a given temperature gradient, even small differences in conductivity can significantly affect the energy transfer rate from the wall to the droplet.

4.1.2 Droplet Impact Dynamics and Heat Transfer Behaviour at 393K and 453K

The impact behaviour of pure n-hexane droplets on surfaces heated to 393K and 453K was analysed to investigate the onset and progression of phase change phenomena as illustrated in Figures 8 and 9, respectively. Initially, the droplet makes gentle contact with the wall and spreads outward due to inertial forces, followed by a brief stationary phase where the droplet maintains a stable, flattened shape. At 393K, nucleate boiling initiates near the droplet–wall interface, leading to the formation of small satellite droplets that gradually rise through the liquid column and detach from the surface [53]. This sequence from initial spreading to satellite droplet ejection is illustrated in Figure 8(a). The observed behaviour suggests that significant heat absorption occurs before the liquid at the wall interface reaches the local boiling point, triggering localised phase change and droplet breakup [53]. At 453K, nucleate boiling becomes more intense, transitioning into a violent regime characterised by rapid vapour bubble formation and energetic ejection of satellite droplets, as shown in Figure 9(a). The increased vapour generation enhances local recoil forces at the interface, resulting in a more aggressive surface breakup. Vapour bubble nucleation within the droplet body is clearly observed, signifying rapid internal heating and strong thermal gradients. This dynamic response highlights the sharp contrast in phase change behaviour with increasing wall temperature and the corresponding enhancement in heat transfer due to vigorous nucleation and recoil.

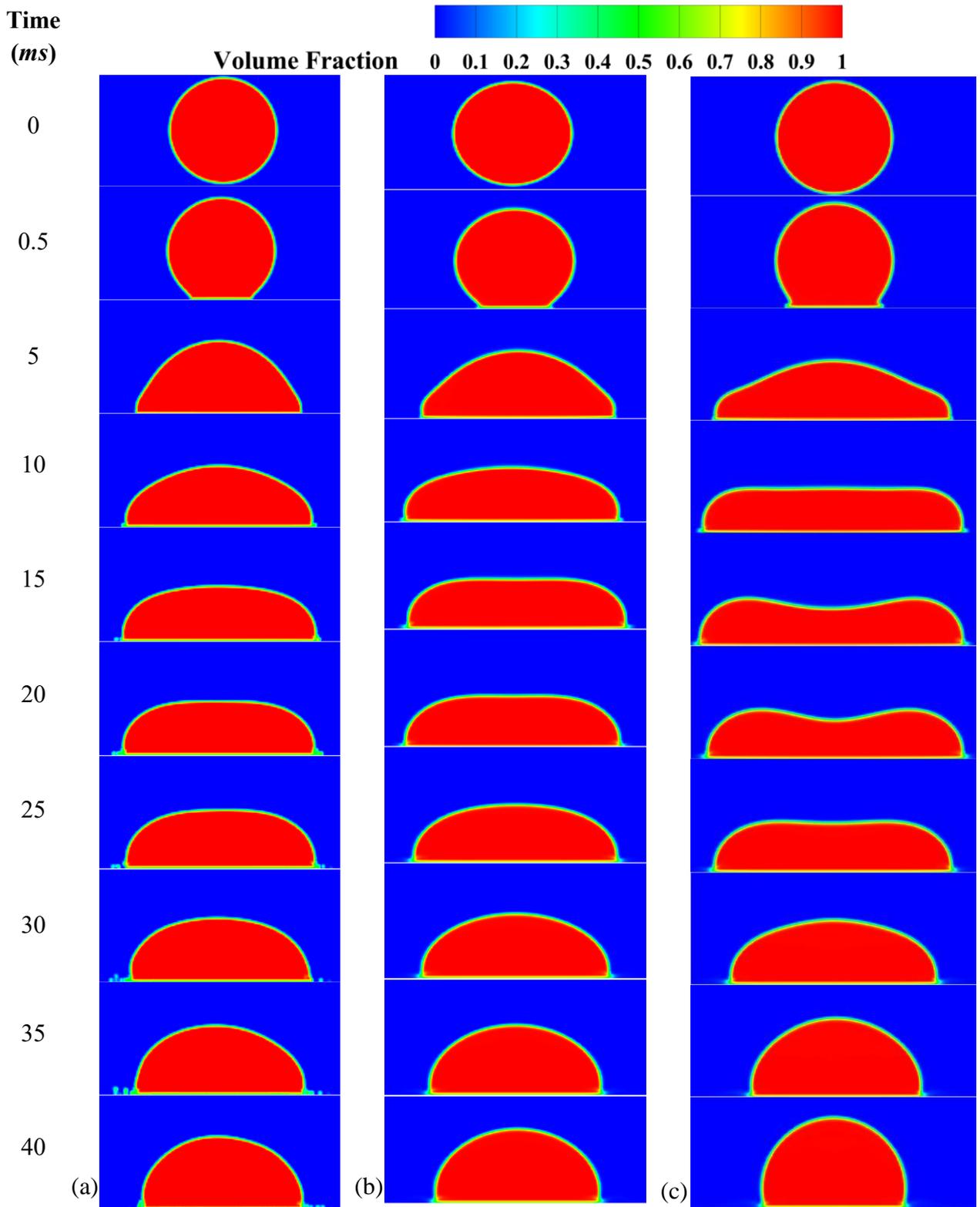

**Figure 8:** Temporal evolution of droplet morphology for (a) pure hexane at Weber number 25, (b) blended fuel (90% Jet A + 10% hexane) at Weber number 25, and (c) blended fuel at Weber number 50 during impact on a stainless-steel surface maintained at 393K.

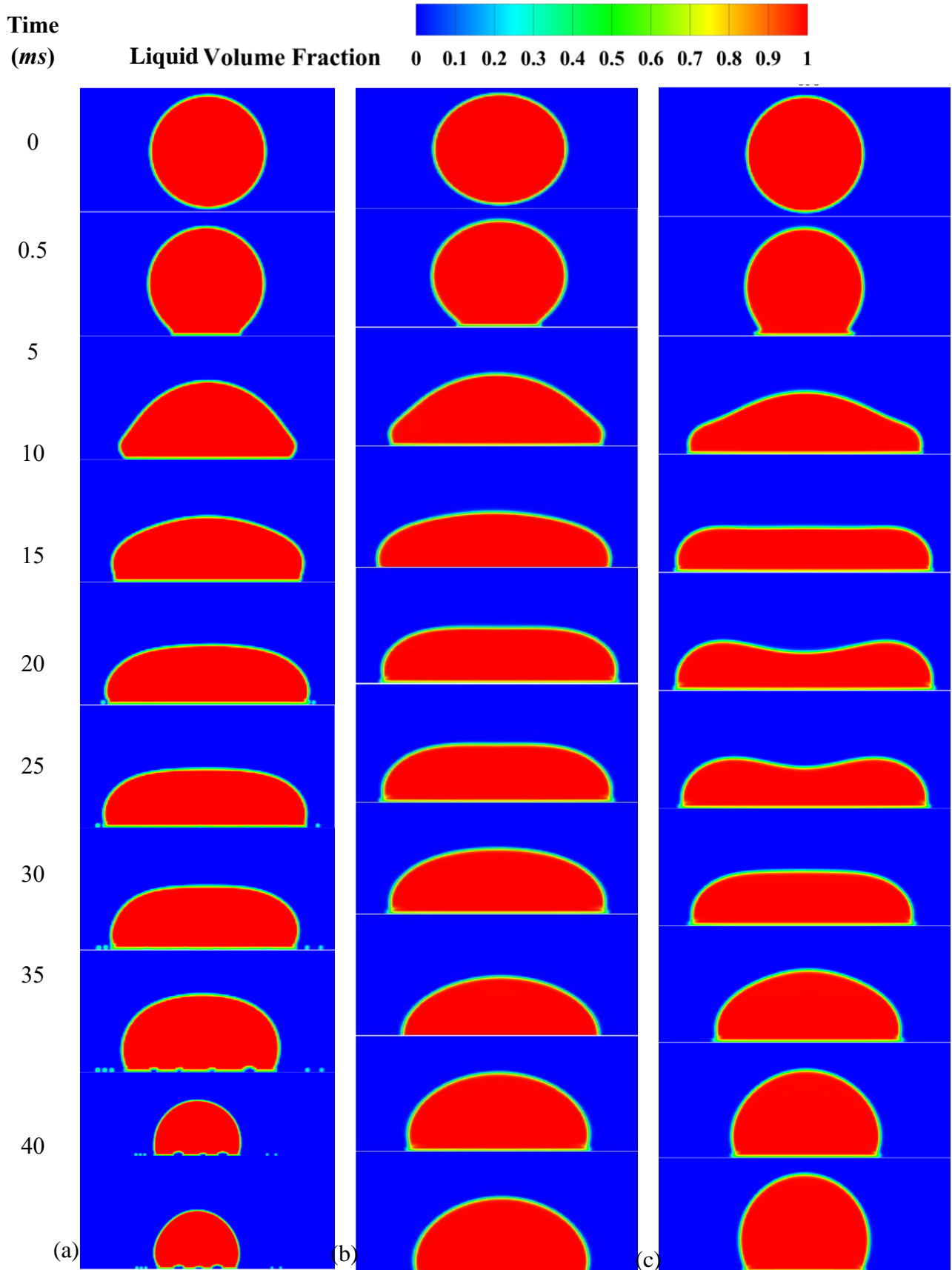

**Figure 9:** Temporal evolution of droplet morphology for (a) pure hexane at Weber number 25, (b) blended fuel (90% Jet A + 10% hexane) at Weber number 25, and (c) blended fuel at Weber number 50 during impact on a stainless-steel surface maintained at 453K.

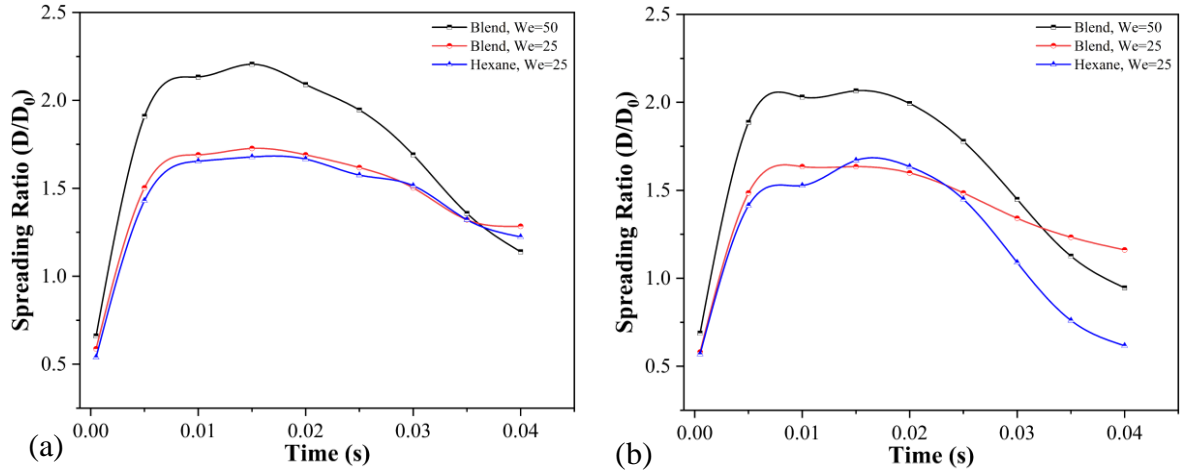

**Figure 10:** Temporal variation of the spreading ratio (D/D$_o$) for pure hexane at Weber number 25 and blended fuel (90% Jet A + 10% hexane) at Weber numbers 25 and 50 during droplet impact on a heated surface at (a) 393K and (b) 453K.

For Blend 25 droplets (90% Jet A + 10% n-hexane) impacting the surface at 393K and 453K, the impact sequence similarly begins with gentle contact and radial spreading. However, unlike n-hexane, the blended droplets exhibit a pronounced receding phase following maximum spread, indicating the dominance of viscous and surface tension forces over inertia during the later stages. As shown in Figure 8(b) and Figure 9(b), no vapour bubbles are observed rising through the droplet interior. Instead, vapour generation is primarily confined to the contact line, suggesting localised evaporation or interfacial boiling rather than bulk nucleation [22]. This suppression of internal boiling is likely due to the higher viscosity and lower thermal conductivity of the blend, as well as the fact that both surface temperatures are below the boiling point of Jet A. Blend 50 droplets (same composition, We = 50) follow a similar spreading and receding sequence, but with key differences arising from their higher impact energy. As shown in Figure 8(c) and Figure 9(c), the higher Weber number promotes lamella formation, a thin liquid sheet spreading radially from the impact centre. The formation of this lamella indicates that increased inertial forces extend the droplet's reach before viscous or capillary forces dominate. Similar to Blend 25, vapour formation is localised near the wall without observable internal bubble nucleation, again pointing to suppressed bulk boiling due to limited thermal penetration and the sub-boiling temperature of Jet A [22].

The temporal variation of the spreading ratio for the three cases at 393K and 453K is presented in Figure 10(a) and Figure 10(b), respectively. At 393K, Blend 50 achieves the highest spreading ratio, driven by its greater inertial energy, followed by Blend 25 and hexane. The gradual decline in the spreading curves reflects mild vapour recoil, which modestly inhibits spreading but does not dominate the dynamics. At 453K, vapour recoil becomes more significant. Hexane shows an early and steep drop in spreading ratio after reaching its peak, highlighting the effect of stronger evaporation-induced forces. In contrast, Blend 25 maintains a more stable spread due to its higher inertia, while Blend 50 still reaches the greatest initial spread but exhibits a sharper receding phase due to the combined effect of high inertia and enhanced evaporation.

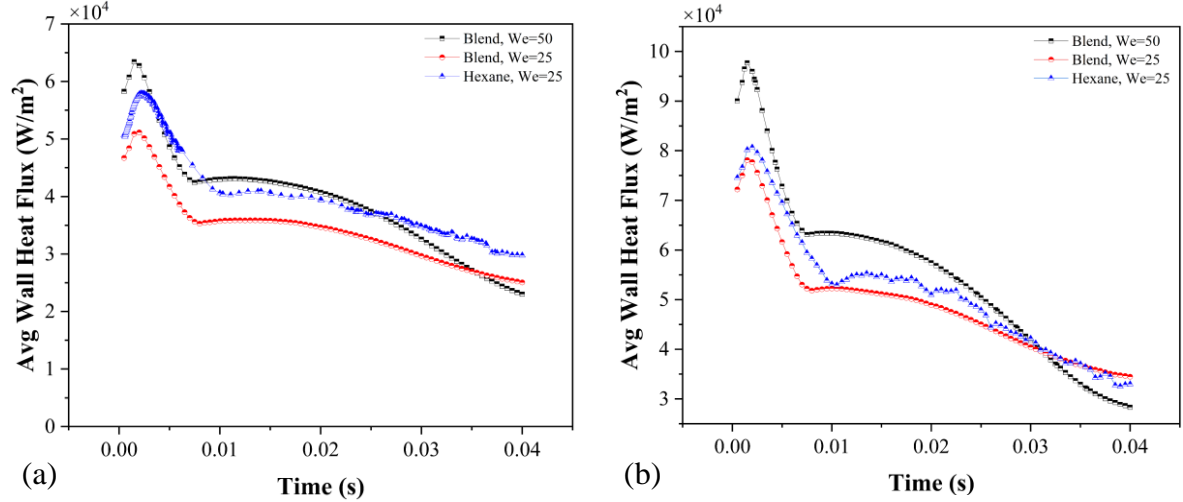

**Figure 11:** Temporal variation of average wall heat flux for pure hexane at Weber number 25 and blended fuel (90% Jet A + 10% hexane) at Weber numbers 25 and 50 during droplet impact on a heated surface at (a) 393K and (b) 453K.

The average wall heat flux at 393K and 453K is shown in Figure 11(a) and Figure 11(b), respectively. Among all cases, Blend 50 exhibits the highest heat flux, primarily due to its extended spreading and larger contact area. Although Blend 25 and hexane have comparable spreading behaviour, hexane exhibits a higher heat flux, which can be attributed to its superior thermal conductivity and specific heat capacity. This demonstrates that when geometric spreading is similar, fluid thermal properties play a dominant role in determining heat transfer efficiency. Figure 12(a) and Figure 12(b) show the wall-adjacent temperature profiles at 0.1 mm above the heated surface. Hexane, with its higher thermal conductivity, receives more heat from the surface; however, due to its higher specific heat capacity, the absorbed heat is more uniformly distributed throughout the droplet, resulting in a lower local temperature rise near the wall. In contrast, Blend 25 with both lower thermal conductivity and specific heat accumulates heat near the contact line, causing a steeper local temperature gradient. Consequently, the wall-adjacent temperature is highest for Blend 50, followed by Blend 25, and lowest for hexane, despite hexane exhibiting the highest average heat flux. These results emphasise the complex interplay between inertia, thermal properties, and phase change behaviour in governing the overall heat transfer and morphological evolution during droplet impact at elevated surface temperatures.

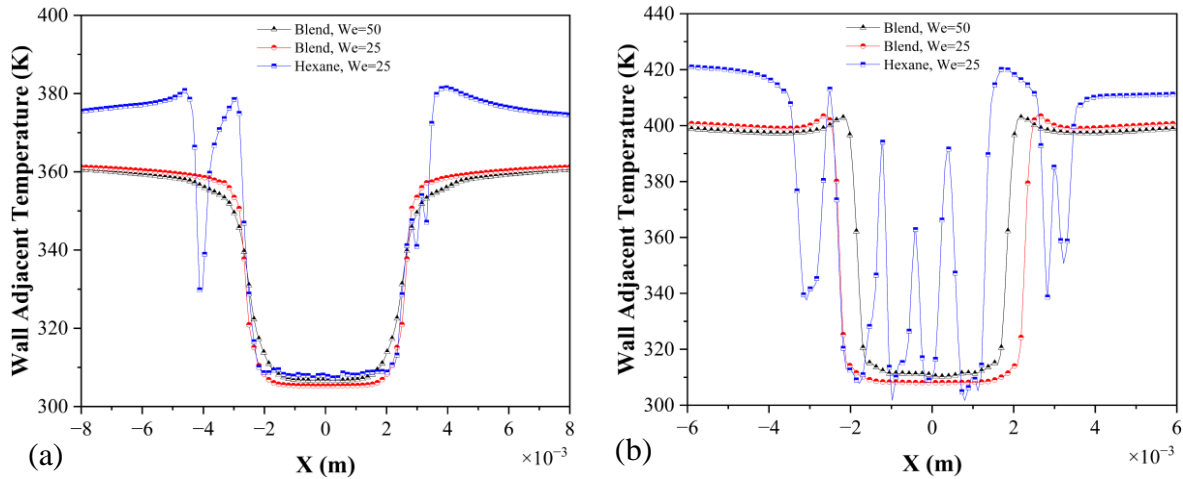

**Figure 12:** Spatial distribution of wall-adjacent temperature at 0.1 mm above the heated surface for pure hexane at Weber number 25 and blended fuel (90% Jet A + 10% hexane) at Weber numbers 25 and 50 during droplet impact on a surface maintained at (a) 393K and (b) 453K.

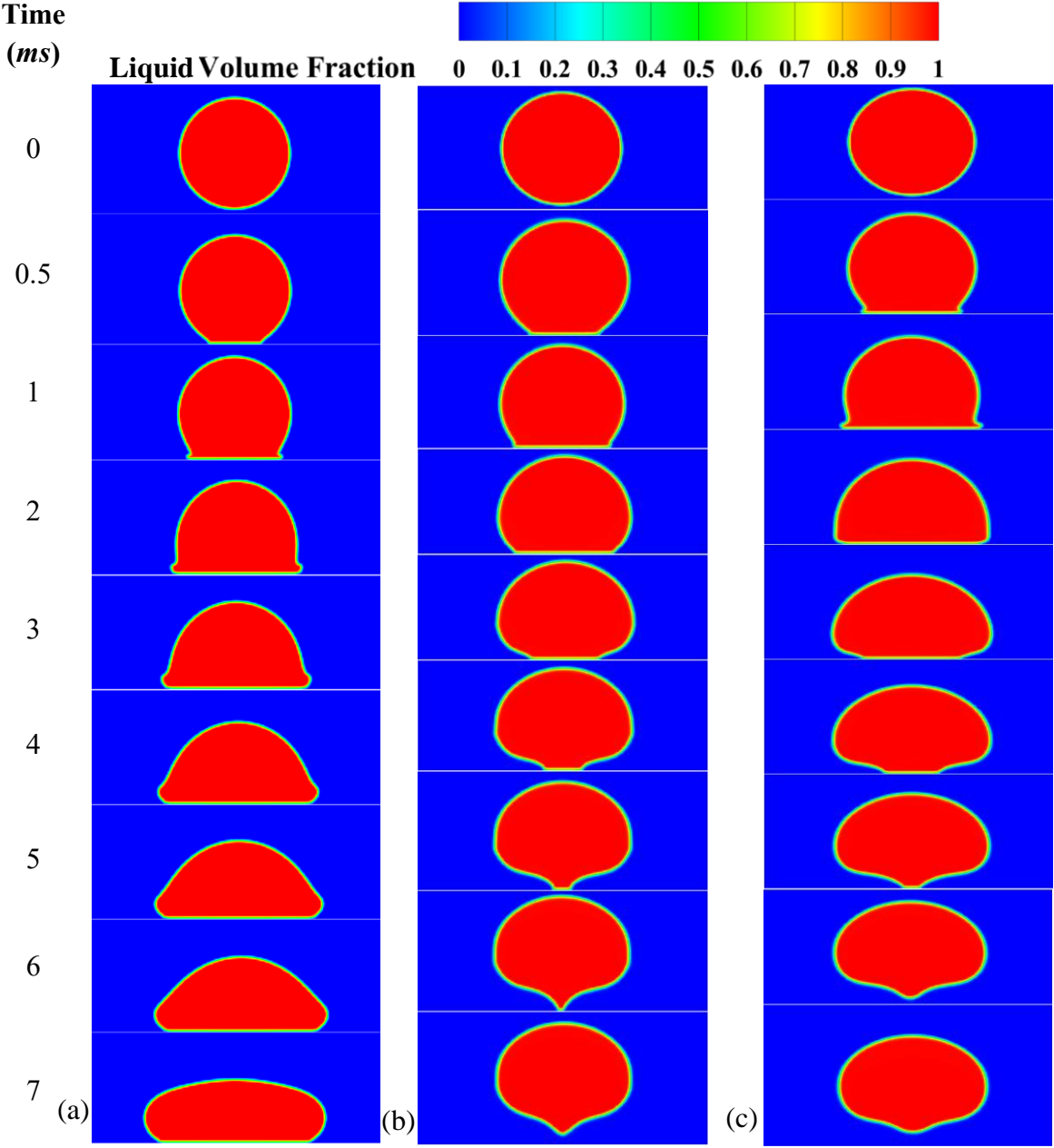

**Figure 13:** Temporal evolution of droplet morphology for (a) pure hexane at Weber number 25, (b) blended fuel (90% Jet A + 10% hexane) at Weber number 25, and (c) blended fuel at Weber number 50 during impact on a stainless-steel surface maintained at 500K.

### 4.1.3 Droplet Impact Dynamics and Heat Transfer Behaviour at 500K

The impact behaviour of Blend 25 and Blend 50 droplets on a stainless-steel surface heated to 500K was investigated to understand droplet wall interactions under extreme thermal conditions as shown in Figure 13. Upon release from a fixed height, each droplet impacted the heated substrate and spread radially due to its initial inertia while simultaneously absorbing heat from the surface. The resulting steep thermal gradient at the liquid-solid interface induced rapid localised vaporisation of both n-hexane and Jet A components. As the droplet spread, vigorous vapor generation beneath the liquid initiated a recoil force normal to the substrate. This vapor recoil gradually intensified and eventually overcame gravitational and adhesive forces, causing the droplet to lift off or bounce from the surface. Prior to lift-off, significant shape distortion was observed, caused by the imbalance between upward recoil and restoring forces such as surface tension and viscosity. The presence of Jet A in the blends amplified this distortion, as its higher viscosity suppressed internal circulation and its lower thermal conductivity facilitated heat retention near the wall, thereby enhancing localized vapor formation.

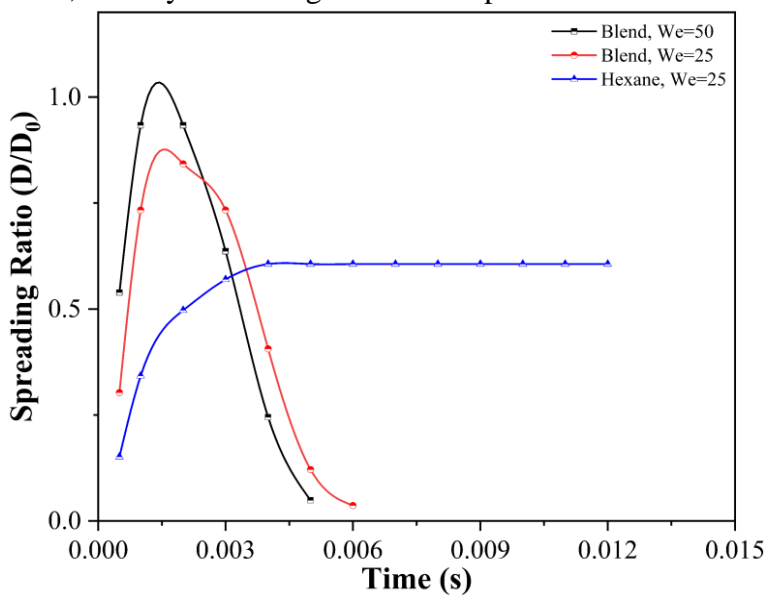

**Figure 14:** Temporal variation of the spreading ratio ($D/D_0$) for pure hexane at Weber number 25 and blended fuel (90% Jet A + 10% hexane) at Weber numbers 25 and 50 during droplet impact on a stainless-steel surface maintained at 500K.

In contrast, the behaviour of pure hexane at 500K differed significantly. Due to the surface temperature being well above its boiling point, the droplet entered a stable film boiling regime. In this regime, a continuous vapour layer formed at the interface, insulating the droplet from the wall and preventing rebound. Instead of bouncing, the droplet remained levitated on the vapor film, exhibiting reduced deformation and prolonged residence without direct contact. The sequence of deformation and rebound dynamics for all cases is illustrated in Figures 13(a), 13(b), and 13(c), highlighting the maximum deformation stage prior to lift-off. The evolution of the spreading ratio for Blend 25 and Blend 50 up to the point of droplet lift-off is shown in Figure 14. Both blends initially exhibit rapid spreading driven by inertia, but their spreading ratios peak sharply and then decline quickly due to the onset of recoil-induced detachment from the surface. This differs from lower temperature regimes, where receding was governed by viscous and capillary forces.

Heat transfer behaviour at 500K further confirms the influence of boiling regime transitions where the average wall heat flux with respect to time is illustrated in Figure 15. As shown in Figure 15(a), Blend 25 and Blend 50 exhibit a steep and rapid decline in average wall heat flux shortly after impact. This behaviour is characteristic of vigorous nucleate boiling followed by droplet rebound. In contrast, hexane displays a smoother and more gradual heat flux decay, consistent with stable film boiling where the vapour layer reduces wall heat flux more progressively. The wall-adjacent temperature distributions, shown in Figure 15(b), provide additional insight. For hexane, the profile reveals a wider and more uniform temperature dip across the droplet footprint, indicating consistent vapour film formation and extended surface coverage. Meanwhile, Blend 25 and Blend 50 show sharper and narrower dips in temperature. These profiles suggest brief contact followed by rebound, resulting in reduced thermal conduction to the wall and higher local wall temperatures compared to hexane. Overall, the results demonstrate that at elevated surface temperatures, the boiling regime strongly governs droplet morphology, contact behaviour, and heat transfer characteristics. While pure hexane undergoes stable film boiling, the blends experience vigorous nucleate boiling and vapour recoil, leading to droplet lift-off. The presence of Jet A intensifies these effects through altered thermal and viscous properties, reinforcing the importance of fuel composition in high-temperature spray–wall interaction scenarios.

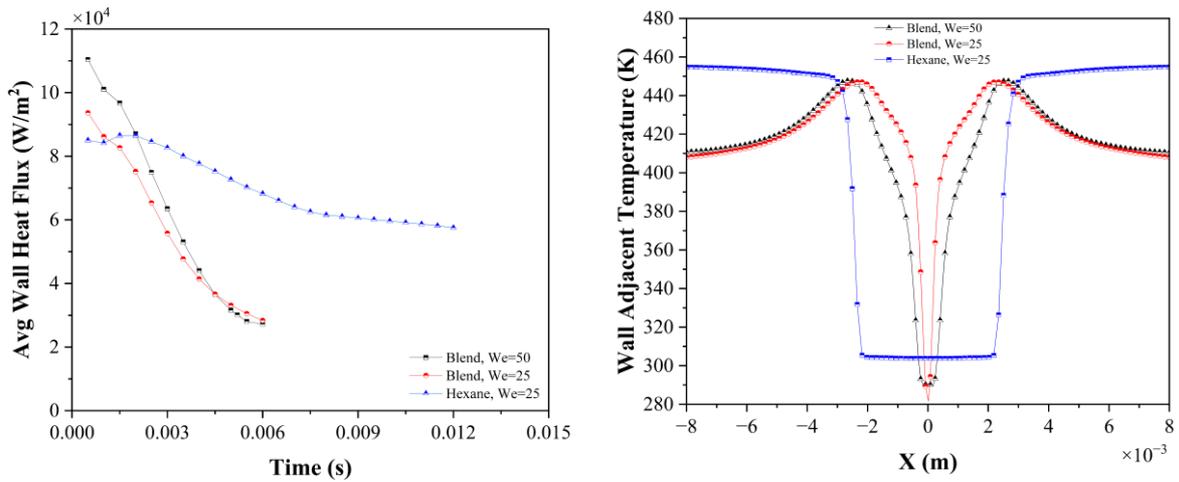

**Figure 15:** (a) Temporal variation of average wall heat flux and (b) spatial distribution of wall-adjacent temperature at 0.1 mm above the surface for pure hexane at Weber number 25 and blended fuel (90% Jet A + 10% hexane) at Weber numbers 25 and 50 during droplet impact on a stainless-steel surface maintained at 500K.

**4.2. Spreading Characteristics Correlation**

To understand the combined influence of Weber number (We), Ohnesorge number (Oh), and surface temperature on the maximum spreading factor ($\beta_{max}$) of blended fuel droplets, empirical correlations were developed under varying thermal conditions. The analysis focuses on the relationship between the maximum spreading factor and the non-dimensional group $We^2 \cdot Oh$, which captures the interplay between inertial, viscous, and capillary effects. Figures 16, 17, 18, 19 show the variation of $\beta_{max}$ with respect to the non-dimensional group $We^2 \cdot Oh$ for three representative cases: pure hexane, Blend 25 (10% hexane + 90% Jet A by volume at We 25), and Blend 50 (10% hexane + 90% Jet A by volume at We 50). The spreading behaviour was systematically evaluated at surface temperatures of 323K, 393K, 453K and 500K.

At 323K, a relatively high spreading factor is observed for all cases. This is primarily attributed to the weak vaporisation at the liquid–solid interface under lower thermal conditions, which allows inertial forces to dominate and promotes broader radial spreading. The empirical correlation derived for this condition, based on curve fitting of the numerical results, is expressed in Equation 22 and visualised in Figure 16:

$$\beta_{max} = 2.2284(Oh.We^2)^{0.1427} \qquad R^2 = 0.9928 \qquad (22)$$

This relation, as illustrated in Figure 16, demonstrates a strong fit to the numerical data, with a coefficient of determination (R²) of 0.9928, indicating a high level of predictive accuracy for spreading behaviour at 323K.

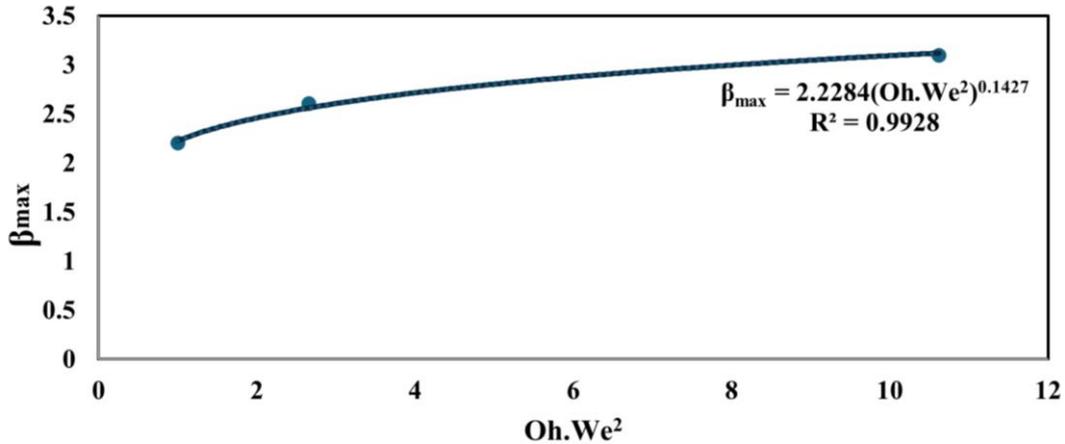

**Figure 16:** Maximum Spreading ratio and Weber number shown by the best-fitted curve for surface temperature 323K.

When the surface temperature is increased to 393K, the maximum spreading factor decreases compared to the 323K condition. This reduction is attributed to the onset of significant vapour generation at the liquid–solids interface, which induces a recoil force that resists the outward motion of the droplet. As a result, the inertial spreading is partially suppressed, leading to a lower maximum spread. The empirical correlation derived for this temperature is given in Equation 23 as:

$$\beta_{max} = 1.6339(Oh.We^2)^{0.1187} \qquad R^2 = 0.9325 \qquad (23)$$

This relation, as shown in Figure 17, captures the trend of reduced spreading under moderate thermal recoil, reflecting the competing effects of inertia and phase change during droplet–wall interaction at 393K.

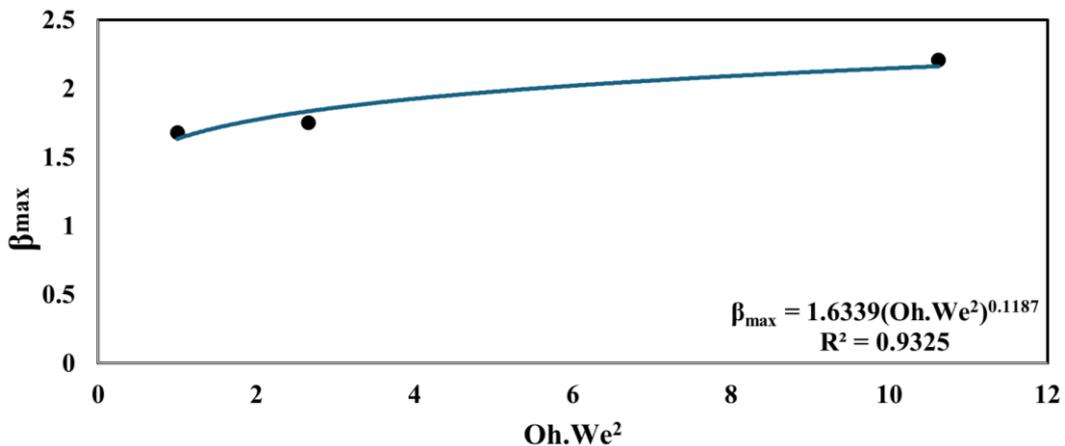

**Figure 17:** Maximum Spreading ratio and Weber number shown by the best-fitted curve for surface temperature 393K.

At a surface temperature of 453K, the maximum spreading factor is further reduced compared to lower temperature conditions. This decline is primarily due to dominant interfacial vapor generation, which introduces substantial upward recoil forces that not only resist radial spreading but also trigger earlier droplet retraction or rebound. The resulting suppression of spreading is more pronounced than at 393K, reflecting the increasing influence of phase change-driven dynamics over inertia. The empirical correlation for this thermal regime is given in Equation 24 as:

$$\beta_{max} = 1.5857(Oh.We^2)^{0.1081} \qquad R^2 = 0.926 \qquad (24)$$

As illustrated in Figure 18, this fitted relation accurately characterises the observed spreading behaviour at 453K, emphasising the role of recoil-induced lift in moderating droplet spreading as surface temperature increases.

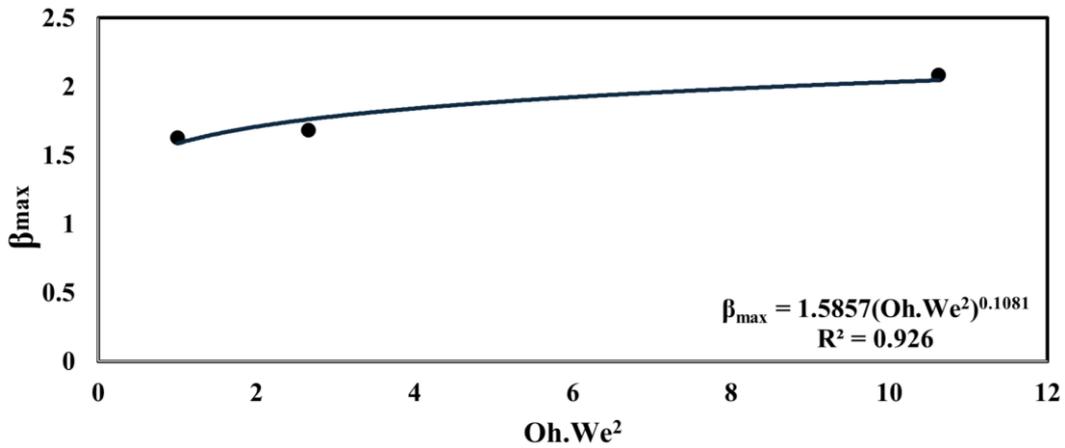

**Figure 18:** Maximum Spreading ratio and Weber number shown by the best-fitted curve for surface temperature 453K.

At a surface temperature of 500K, the maximum spreading factor decreases further compared to lower temperature cases. In this regime, distinct differences in boiling behaviour were observed among the fuel types. Pure hexane droplets underwent stable film boiling, characterised by the formation of a continuous vapour layer that insulated the droplet from the surface and suppressed rebound. In contrast, the Jet A component alone predominantly exhibited nucleate boiling, with localised vapour bubble formation but insufficient recoil to cause lift-off. For the blended fuel droplets, the combination of highly volatile hexane and the thermally retentive Jet A led to intense interfacial vapour generation. This strong vapour production generated significant recoil forces beneath the droplet, causing shape distortion and, ultimately, droplet rebound despite the high substrate temperature. The empirical correlation as given in Equation 25, capturing this spreading behaviour at 500K, is:

$$\beta_{max} = 0.6223(Oh.We^2)^{0.2638} \qquad R^2 = 0.9863 \qquad (25)$$

As shown in Figure 19, this relation provides an excellent fit to the numerical data, with a high coefficient of determination indicating robust predictive accuracy.

Overall, these findings highlight the dual influence of inertial and thermal effects on droplet spreading behavior. While an increase in inertial forces (via higher Weber number) typically enhances radial spreading, the influence of surface temperature is equally significant. At elevated temperatures, enhanced vapor generation at the contact line introduces resistive recoil forces that substantially reduce the maximum spreading factor. The strong agreement between simulation results and the proposed empirical relations across all thermal conditions confirms the validity and utility of the developed correlations for predicting droplet spreading dynamics of blended fuels in high-temperature environments.

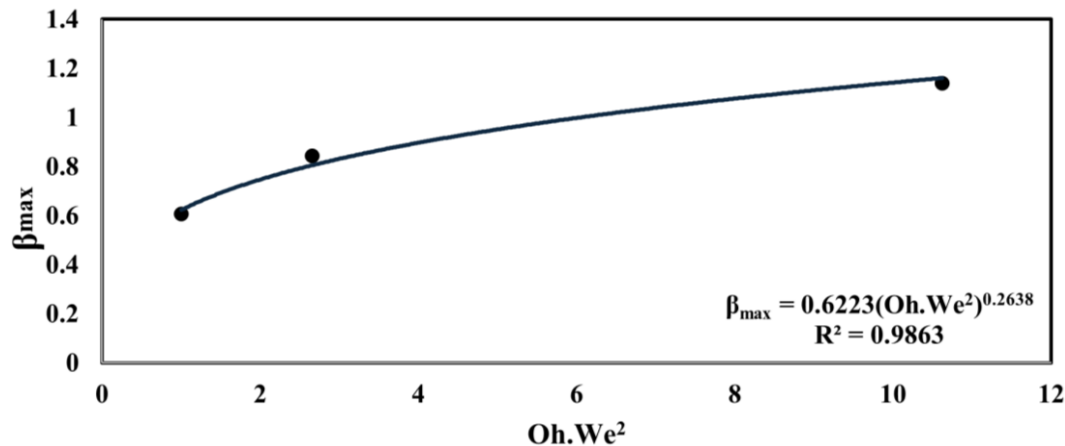

**Figure 19:** Maximum Spreading ratio and Weber number shown by the best-fitted curve for surface temperature 500K.

## 5. CONCLUSIONS

This work evaluated the impact of a single blended fuel droplet (comprising 90% Jet A and 10% n-hexane by volume) on a heated stainless-steel surface under varying thermal and inertial conditions, with a focus on characterising droplet dynamics and heat transfer behaviour. Numerical simulations were performed using the Volume of Fluid (VOF) approach, where a user-defined function (UDF) was implemented to model the velocity-dependent dynamic contact angle at the triple-phase contact line. This addition ensured accurate representation of wettability and spreading phenomena during transient droplet deformation. This approach systematically examines the surface temperature, fuel composition, and Weber number that govern spreading, recoil, and thermal interactions during impact.

The impact dynamics and heat transfer behaviour of blended fuel droplets were evaluated across four surface temperatures: 323K, 393K, 453K and 500K. At 323K, spreading was primarily governed by inertia, with Blend 25 spreading more than hexane due to its higher density, and Blend 50 achieving the greatest spread from higher impact velocity. Hexane, despite spreading less, exhibited the highest heat flux due to its superior thermal conductivity. As temperature increased to 393K and 453K, phase change effects intensified. Hexane underwent nucleate and violent boiling, limiting spreading through vapour recoil, while Blend 25, constrained by higher viscosity, showed localised vaporisation and receded smoothly. Blend 50's greater inertia led to extended spreading and lamella formation before recoil took over. At 500K, hexane entered a film boiling regime with vapour layer formation and no rebound, whereas Blend 25 and Blend 50 experienced vigorous interfacial vaporisation, strong recoil, and droplet lift-off.

Empirical correlations were developed using the combined non-dimensional group $We^2 \cdot Oh$ to predict the maximum spreading factor for each thermal condition, with strong agreement observed ($R^2 > 0.92$). The implementation of a user-defined function (UDF) to model the dynamic contact angle, dependent on contact line velocity, further enhanced the accuracy of droplet morphology prediction. While the study offers valuable insights into droplet–surface interaction, it is limited to single-droplet impacts on smooth, isothermal surfaces without accounting for surface roughness, spray interactions, or transient heating. Future work should explore complex boundary conditions, multicomponent fuel systems, and experimental validation to extend the applicability of the findings to real-world thermal management and spray combustion scenarios.

## Acknowledgements

The authors would like to acknowledge the Department of Science and Technology, Government of India, for financial support for this research (Grant no. CRG/2022/007787).

## References

[1]     Chandra S, Avedisian CT. On the collision of a droplet with a solid surface. Proc R Soc A Math Phys Eng Sci 1991;432:13–41. https://doi.org/10.1098/rspa.1991.0002.

[2]     Moita AS, Moreira ALN. Development of empirical correlations to predict the secondary droplet size of impacting droplets onto heated surfaces. Exp Fluids 2009;47:755–68. https://doi.org/10.1007/s00348-009-0719-1.

[3]     Cossali GE, Marengo M, Santini M. Single-drop empirical models for spray impact on solid walls: A review. At Sprays 2005;15:699–736. https://doi.org/10.1615/AtomizSpr.v15.i6.50.

[4]     Moreira ALN, Moita AS, Panão MR. Advances and challenges in explaining fuel spray impingement: How much of single droplet impact research is useful? Prog Energy Combust Sci 2010;36:554–80. https://doi.org/10.1016/j.pecs.2010.01.002.

[5]     Raj K, Lawrence Raj PR. Investigation of Dynamic Characteristics of Liquid Methane Droplet Impact on a Heated Solid Surface. Heat Transf Res 2025;56:49–67. https://doi.org/10.1615/HeatTransRes.2025057349.

[6]     Bai CX, Rusche H, Gosman AD. MODELING OF GASOLINE SPRAY IMPINGEMENT. At Sprays 2002;12:1–28. https://doi.org/10.1615/AtomizSpr.v12.i123.10.

[7]     AAS 1003-05 (387-408).pdf n.d.

[8]     Sivakumar D, Tropea C. Splashing impact of a spray onto a liquid film. Phys Fluids 2002;14:L85–8. https://doi.org/10.1063/1.1521418.

[9]     Kim J. Spray cooling heat transfer: The state of the art. Int J Heat Fluid Flow 2007;28:753–67. https://doi.org/10.1016/j.ijheatfluidflow.2006.09.003.

[10]    Mühlbauer M, Roisman I V., Tropea C. Evaluation of spray/wall interaction data. Meas Sci Technol 2011;22. https://doi.org/10.1088/0957-0233/22/6/065402.

[11]    Karl A, Frohn A. Experimental investigation of interaction processes between droplets and hot walls. Phys Fluids 2000;12:785–96. https://doi.org/10.1063/1.870335.

[12]    Kang BS, Lee DH. On the dynamic behavior of a liquid droplet impacting upon an inclined heated surface. Exp Fluids 2000;29:380–7.

https://doi.org/10.1007/s003489900104.

[13] Cossali GE, Marengo M, Santini M. Thermally induced secondary drop atomisation by single drop impact onto heated surfaces. Int J Heat Fluid Flow 2008;29:167–77. https://doi.org/10.1016/j.ijheatfluidflow.2007.09.006.

[14] Tran T, Staat HJJ, Prosperetti A, Sun C, Lohse D. Drop Impact on Superheated Surfaces. Phys Rev Lett 2012;108:036101. https://doi.org/10.1103/PhysRevLett.108.036101.

[15] Crafton EF, Black WZ. Heat transfer and evaporation rates of small liquid droplets on heated horizontal surfaces. Int J Heat Mass Transf 2004;47:1187–200. https://doi.org/10.1016/j.ijheatmasstransfer.2003.09.006.

[16] Bhardwaj R, Longtin JP, Attinger D. Interfacial temperature measurements, high-speed visualization and finite-element simulations of droplet impact and evaporation on a solid surface. Int J Heat Mass Transf 2010;53:3733–44. https://doi.org/10.1016/j.ijheatmasstransfer.2010.04.024.

[17] Chatzikyriakou D, Walker SP, Hale CP, Hewitt GF. The measurement of heat transfer from hot surfaces to non-wetting droplets. Int J Heat Mass Transf 2011;54:1432–40. https://doi.org/10.1016/j.ijheatmasstransfer.2010.11.051.

[18] Bhat M, Sakthikumar R, Sivakumar D. Fuel drop impact on heated solid surface in film evaporation regime. Chem Eng Sci 2019;202:95–104. https://doi.org/10.1016/j.ces.2019.03.017.

[19] Sen S, Vaikuntanathan V, Sivakumar D. Impact dynamics of alternative jet fuel drops on heated stainless steel surface. Int J Therm Sci 2017;121:99–110. https://doi.org/10.1016/j.ijthermalsci.2017.07.006.

[20] Zhang L, Ma H, Shen Z, Wang L, Liu R, Pan J. Influence of pressure and temperature on explosion characteristics of n-hexane/air mixtures. Exp Therm Fluid Sci 2019;102:52–60. https://doi.org/10.1016/j.expthermflusci.2018.11.004.

[21] Boettcher PA, Mével R, Thomas V, Shepherd JE. The effect of heating rates on low temperature hexane air combustion. Fuel 2012;96:392–403. https://doi.org/10.1016/j.fuel.2011.12.044.

[22] Kompinsky E, Dolan G, Sher E. Experimental study on the dynamics of binary fuel droplet impacts on a heated surface. Chem Eng Sci 2013;98:186–94. https://doi.org/10.1016/j.ces.2013.04.047.

[23] Kitano T, Nishio J, Kurose R, Komori S. Evaporation and combustion of multicomponent fuel droplets. Fuel 2014;136:219–25. https://doi.org/10.1016/j.fuel.2014.07.045.

[24] Chausalkar A, Kong SC, Michael JB. Multicomponent drop breakup during impact with heated walls. Int J Heat Mass Transf 2019;141:685–95. https://doi.org/10.1016/j.ijheatmasstransfer.2019.06.033.

[25] Nikolopoulos N, Theodorakakos A, Bergeles G. A numerical investigation of the evaporation process of a liquid droplet impinging onto a hot substrate. Int J Heat Mass Transf 2007;50:303–19. https://doi.org/10.1016/j.ijheatmasstransfer.2006.06.012.

[26] Strotos G, Gavaises M, Theodorakakos A, Bergeles G. Numerical investigation of the evaporation of two-component droplets. Fuel 2011;90:1492–507. https://doi.org/10.1016/j.fuel.2011.01.017.

[27] Banerjee R. Numerical investigation of evaporation of a single ethanol/iso-octane droplet. Fuel 2013;107:724–39. https://doi.org/10.1016/j.fuel.2013.01.003.

[28] Huang R, Gong J, Jiang Z, Zhang N, Hou J. Numerical simulation study of droplet impact on microscale groove-textured superhydrophobic surface: Influence of geometric structural parameters. Colloids Surfaces A Physicochem Eng Asp 2025;709:136053. https://doi.org/10.1016/j.colsurfa.2024.136053.

[29] Wang Q, Liu S, Wang W, Zhao H, Zhang X, Han F, et al. Dynamic characteristics of droplets' impact on solid surfaces with varied roughness. Phys Fluids 2025;37. https://doi.org/10.1063/5.0249633.

[30] Wang M, Wang F, Wang X, Yang B, Yu W. The effect of tin droplet impact velocity and stainless steel substrate temperature on droplet splashing behavior. Appl Phys A Mater Sci Process 2025;131. https://doi.org/10.1007/s00339-025-08300-9.

[31] Azizifar S, Song M, Chao CYH, Hosseini SH, Pekař L. A numerical study of multiphase flow boiling heat transfer of nanofluids in the horizontal metal foam tubes. Int J Thermofluids 2024;22. https://doi.org/10.1016/j.ijft.2024.100605.

[32] Tao J, Jin J, He X, Lin Z, Zhu Z. Heat transfer of symmetric impacts of two droplets on a hot liquid film. Chem Eng Sci 2024;299:120516. https://doi.org/10.1016/j.ces.2024.120516.

[33] Tao J, Zhu H, Chen D, Lin Z, Zhu Z. Heat transfer properties of a droplet colliding with a liquid film on a protruding surface. Int J Heat Mass Transf 2025;239:126574. https://doi.org/10.1016/j.ijheatmasstransfer.2024.126574.

[34] James WO, Perez-Raya I. Three-Dimensional Simulations of Nucleate Boiling With Sharp Interface Volume of Fluid and Localized Adaptive Mesh Refinement in ANSYS-FLUENT. ASME J Heat Mass Transf 2024;146:1–11. https://doi.org/10.1115/1.4064459.

[35] Linstrom PJ, Mallard WG. The NIST Chemistry WebBook: A chemical data resource on the internet. J Chem \& Eng Data 2001;46:1059–63.

[36] Yu J, Eser S. Determination of Critical Properties (Tc, 1996:404–9.

[37] Mishra S, Bukkarapu KR, Krishnasamy A. A composition based approach to predict density, viscosity and surface tension of biodiesel fuels. Fuel 2021;285:119056. https://doi.org/10.1016/j.fuel.2020.119056.

[38] Ramírez-Verduzco LF. Models for predicting the surface tension of biodiesel and methyl esters. Renew Sustain Energy Rev 2015;41:202–16. https://doi.org/10.1016/j.rser.2014.08.048.

[39] Fydrych D, Rogalski G. Effect of underwater local cavity welding method conditions on diffusible hydrogen content in deposited metal. Weld Int 2013;27:196–202. https://doi.org/10.1080/09507116.2011.600033.

[40] LAUNDER BE, SPALDING DB. the Numerical Computation of Turbulent Flows. vol. 3. Pergamon Press, Ltd; 1983. https://doi.org/10.1016/b978-0-08-030937-8.50016-7.

[41] de Almeida E, Spogis N, Taranto OP, Silva MA. Theoretical study of pneumatic separation of sugarcane bagasse particles. Biomass and Bioenergy 2019;127:105256.

https://doi.org/10.1016/j.biombioe.2019.105256.

[42]  Shah F, Fall I, Zhang D. Experimental and CFD evaluation of bubble diameter and turbulence model influence on nonlinear flow dynamics in vertical columns: A comparative study. Chaos, Solitons and Fractals 2025;196:116421. https://doi.org/10.1016/j.chaos.2025.116421.

[43]  Chapter I. Introduction. A Theor. Study Interphase Mass Transf., Columbia University Press; 1953, p. 1–24. https://doi.org/10.7312/schr90162-001.

[44]  Berg J, editor. Wettability. CRC Press; 1993. https://doi.org/10.1201/9781482277500.

[45]  Os COS, Om COS. 【Jiang1979】Correlation for Dynamic Contact Angel.Pdf 1979;69.

[46]  Bracke M, Voeght F, Joos P. The kinetics of wetting: the dynamic contact angle. Trends Colloid Interface Sci III 2007;149:142–9. https://doi.org/10.1007/bfb0116200.

[47]  Seebergh JE, Berg JC. Dynamic wetting in the low capillary number regime. Chem Eng Sci 1992;47:4455–64. https://doi.org/10.1016/0009-2509(92)85123-S.

[48]  Cox RG. The dynamics of the spreading of liquids on a solid surface. Part 2. Surfactants. J Fluid Mech 1986;168:195–220. https://doi.org/10.1017/S0022112086000344.

[49]  Miller C. Liquid Water Dynamics in a Model Polymer Electrolyte Fuel Cell Flow Channel by. Thesis 2007:128.

[50]  Deendarlianto, Pradecta MR, Prakoso T, Indarto, Mitrakusuma WH, Widyaparaga A. Contact angle dynamics during the impact of single water droplet onto a hot flat practical stainless steel surface under medium Weber numbers. Heat Mass Transf Und Stoffueberagung 2021;57:1097–106. https://doi.org/10.1007/s00231-020-03010-9.

[51]  Mehdi-Nejad V, Mostaghimi J, Chandra S. Air bubble entrapment under an impacting droplet. Phys Fluids 2003;15:173–83. https://doi.org/10.1063/1.1527044.

[52]  Pradhan TK, Panigrahi PK. Thermocapillary convection inside a stationary sessile water droplet on a horizontal surface with an imposed temperature gradient. Exp Fluids 2015;56:1–11. https://doi.org/10.1007/s00348-015-2051-2.

[53]  Mitrakusuma WH, Deendarlianto, Kamal S, Indarto, Nuriyadi M. Experimental investigation on the phenomena around the onset nucleate boiling during the impacting of a droplet on the hot surface. AIP Conf Proc 2016;1737. https://doi.org/10.1063/1.4949305.